# Improving Performance of Routing Protocols Using MRP Framework

Sohail Abid[1]
Department of Engineering and Management
Foundation University, Pakistan.
rsohailabid@yahoo.com

Shahab Khan[2]
Department of Computing and Technology
IQRA University Islamabad Campus Pakistan.
shahabiqra@gmail.com

*Abstract*—These days MANET (Mobile Ad-hoc Network) is an amazing remarkably altering or rising technology, for the reason that of its elite nature of scattered mobile devices and self-motivated network topology. The mobile ad-hoc routing protocol follows several principles in wireless MANET's. The up to date and novel applications based on wireless technology are being produced in the private as well as commercial sectors. A lot of challenges which are facing wireless MANETs like network stability, security, energy efficiency and performance analysis etc. At present wireless ad-hoc network get much more attention because of its accessibility everywhere. As a result researchers produce several routing protocols. In this paper first of all we analyzed the performance investigation of wireless routing protocols on the basis of ROH (Routing Overhead), throughput, end-2-end delay and PDR (Packet Delivery Ratio). After that we proposed an MRP (Mixed Routing Protocol) framework which improve performance.

I. INTRODUCTION

Now a day mobile ad-hoc protocol acting an essential part in a wireless atmosphere. Today mobile network has become a primary element of recent communication infrastructure for its applications in mobile and personal communications. The strength of mobile ad-hoc technology is that the mobile devices can be used anywhere and at any time. In mobile ad-hoc network, all devices work as a router or end node, which participate an significant function during safeguarding and searching of routes. The breakdown of a mobile device can critically modify the performance of an ad - hoc network.

MANET is a collection of wireless devices that set up the relationship between wireless nodes exclusive of centralized management and infrastructure [1]. The wireless nodes are proficient of shifting their location and connect each other randomly in a wireless network. The whole procedure replicates, during finding the whole route, the destination mobile device sends route reply message to the source mobile device for successful route making and searching procedure [2]. The arrangement of route detection and preservation is the important process functioning in DSR [3]. The proactive routing protocol Bellman-Ford method working in DSDV[4]. The protection and marking of routes ought to be finished below some limitations for example utilization of bandwidth and minimum quantity of overhead [5]. The main goal of a routing protocol is to set up a accurate and competent route linking two mobile nodes that can be sent or received in time.

**1.1 MANET Routing Protocols**

Three main categories of routing protocols, which are as under

1. Proactive

2. Reactive

3. Hybrid

**1.2 MANET Networking**

Now a day Mobile Ad-hoc networks have much more awareness, because building of mobile ad-hoc network is very easy or without any requirement of pre-existing infrastructure. A collection of nodes in a wireless ad-hoc network, that are agree and capable to set up without any infrastructure and central supervision. Wireless MANET provides an environment, where each and every mobile node act as a router for example each node receives acknowledgement and forward this acknowledgement to the next hop, in this way the acknowledgement arrive at last mobile node or the destination node from various hops.

**1.3 Mobility Models**

There are many mobility models, but we discuss some of them.

**a. Random Direction**

In random direction mobility model each node chooses a random direction and velocity from the given range and move to the designation and repeat this process till simulation ends.

**b. Random Waypoint**

In random direction mobility model each node chooses a random direction and velocity from the given range and move to the designation and stops for the give pause time and repeat this process till simulation ends.

**c. RPGM (Reference Point Group Mobility Model)**

In RPGM mobility model each node chooses its group and each group has a group leader. The group leader In random direction mobility model each node has random direction and velocity from the given range and move to the designation and stops for the give pause time and repeat this process till simulation ends.

## II. PREVIOUS WORK

Birdar and his group menmers offered in his paper a performance analysis for DSR and AODV using speed as a variable parameter in NS2 [6]. Another performance analysis for DSR and AODV in NS2 using number of sources, speed and pause time as a variable parameter by G. Jayakumar and his group members [7]. Yogesh and his team presented a comparison analysis for DSR and AODV using variable parameters as the number of nodes, speed and pause time in GLOMOSIM [8]. Shaily and his team members compare ZRP, DSR and AODV using pause time as a variable parameter in QualNet [9]. Vijayalaskhmi and his group analyze performance of AODV and DSDV in his paper using pause time as variable parametert in NS2 [10]. Amr M. Hassan and his team select DSDV and DSR and evaluate the performance on the basis of Routing Overhead (ROH), Packet Delivery Latency and Packet Delivery Function (PDF) versus node density [11]. The comparative study of DSDV, AODV and DSR on the basis of "ROH", "Packet Dropped Ratio" and "Packet Delay" by Kaushik et al. [12]. In [13], W.R. S. Jeyaseelan and Sh. Hariharanthe evaluate performance of DSR, OLSR and AODV. Under high mobility Shah and his group members simulate AODV, DSDV and DSR in NS-2 [14]. A Behavioral Study of Routing Protocols like AODV, DSDV and DSR in ns2 by Kumar Sharma and his group [15]. Maashri and his group members perform another comparative study of routing protocols at high mobility [16]. Kumari et al. also investigate performance analysis of OLSR, AODV and DSDV in Freeway mobility model [17].

## III. SIMULATIONS

In this simulation we select four routing protocols, RPGM Mobility model (Reference Point Group Mobility) and node speed from 0.5 to 5.0 m/s, traffic generator source and transport protocols is CBR and UDP. The simulation area is 600 x 600 meter, the propagation model is two-ray and node density change from 20 to 80. Further details are in table 1.

| Simulation Parameters | |
|---|---|
| Transmission range | 600m x 600 m |
| Antenna | Omni directional |
| MAC Type | IEEE 802.11 |
| Node speed | 0.5m/s to 5.0 m/s |
| Simulation Time | 100 sec |
| Packet rate | 8 packets per sec. |
| Traffic Type | CBR |
| Data payload | 512 bytes/ packet |
| Interface Queue Type | Drop Tail/Priori Queue |
| Node Pause Time | 0 |
| Mobility Model | RPGM |
| Interface Queue Length | 50 |
| No. of Nodes | 20 up to 80 |

**Table 1**

**Performance Metric**

The performance metric of routing protocols is described as below.

A. *ROH (Routing Overhead)*

ROH is defined as the total number of routing packets during communication in a simulation.

$$R = \text{No of RTR.}$$

Where "No. of RTR" represents to total number of routing packets and "R" represents Routing Overhead.

B. *PDF (Packet Delivery Function)*

PDF is defined as the total number of sending packets dividing by the total number of received packets.

$$P = Pkt\_S / Pkt\_R$$

Where "Pkt_S" represent send packet and "Pkt_R" represent receive packet and "P" represent packet delivery function.

C. *Average End 2 End Delay (Average E-2-E Delay)*

During communication there are various issues that cause delay like retransmission, queuing, buffering during routes detection and latency. Minimum delay means good performance.

$$\text{E2E Delay} = \frac{1}{m} \sum_{a=1}^{m} (R_a - S_a)$$

$a = \{1, 2, \ldots, m\}$
$R_a$ = time "a" when packet was received by destination
$S_a$ = time "a" when packet was send by source
$m$ = number of receiving packets.

## IV. PROPOSED FRAME WORK AND RESULTS

I propose a new routing protocol framework MRP (Mixed Routing Protocol) on the basis of my results. The framework is as under:

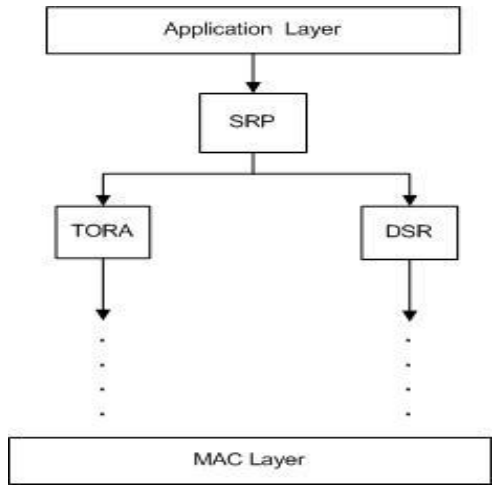

Figure 1a: Proposed MRP

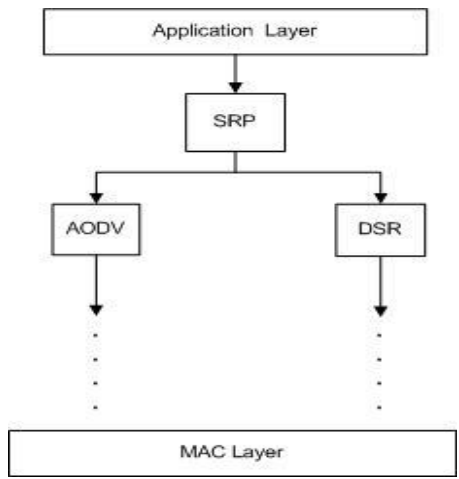

Figure 1b: Proposed MRP

In the framework SRP stands for swap routing protocol. In our proposed framework I select two best routing protocols on the basis of our simulation and SRP switch routing protocols during communication. This method work in application layer. First of all I take two protocols AODV and DSR and simulate in ns2 and compare with other routing protocols. Secondly I change the combination of routing protocols, this time I take Tora and DSR. Simulate in NS2 and compare with other routing protocols. Our MRP framework performs better than other routing protocols.

In other words performance of my proposed framework increased, my new results are as under.

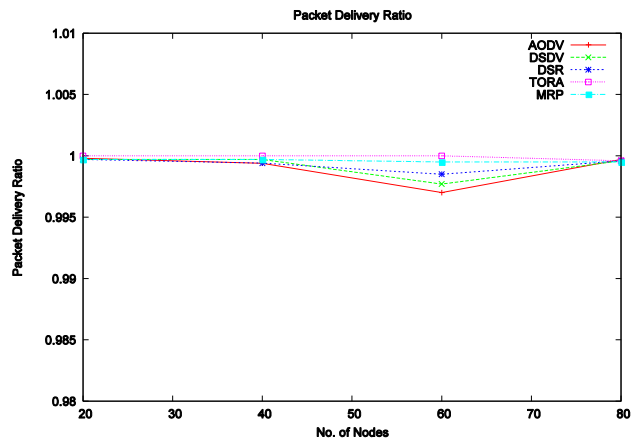

Figure 2: PDF

In the above figure 2, it is clear that our proposed frame gives performance nearly equal to TORA. In other words MRP gives best performance in PDF.

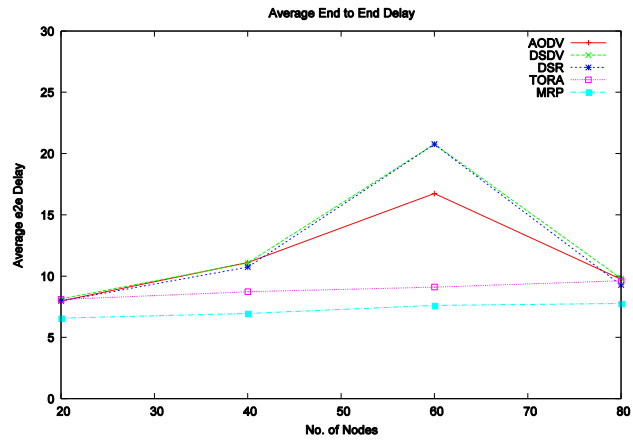

Figure 3: Avg. E2E Delay

In the above figure 3, it is clear that our proposed frame gives performance better than TORA. In other words MRP gives best performance in e2e delay.

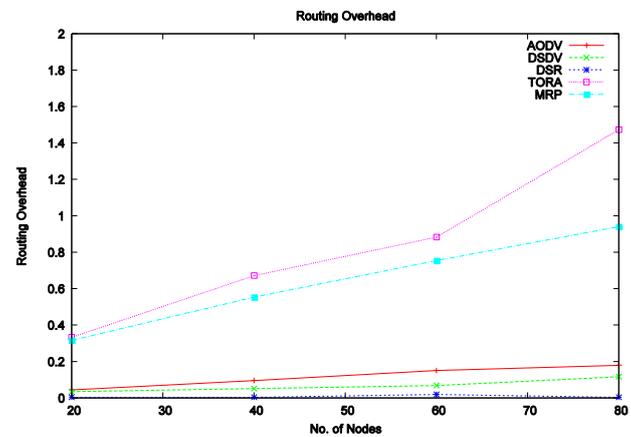

Figure 4: RO

In the above figure 4, it is clear that our proposed frame gives performance nearly equal to DSR. In other words MRP gives best performance in ROH.

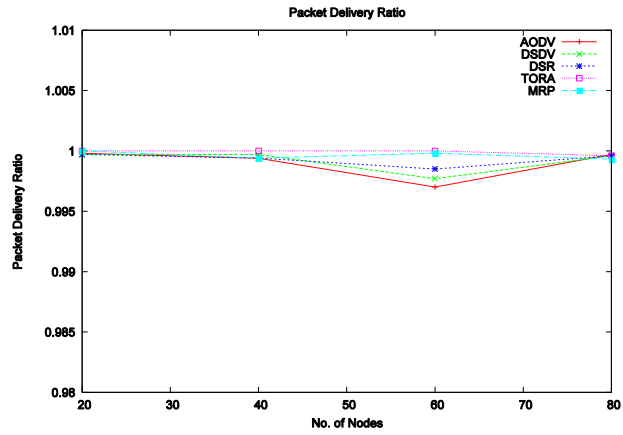

Figure 5: PDF

In the above figure 5, it is clear that our proposed frame gives performance nearly equal to TORA. In other words MRP gives best performance in PDF.

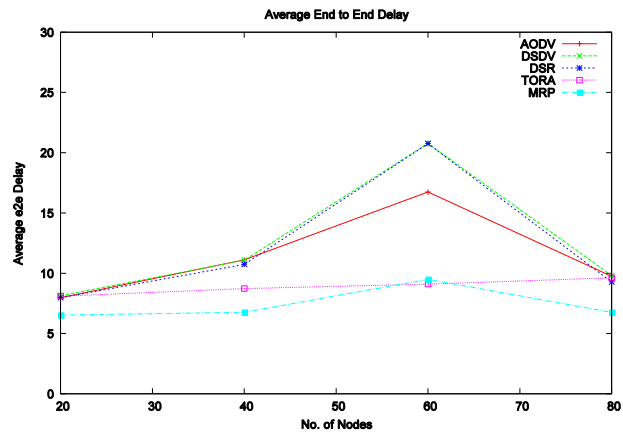

Figure 6: Avg. E2E Delay

In the above figure 6, it is clear that our proposed frame gives performance better than TORA. In other words MRP gives best performance in e2e delay.

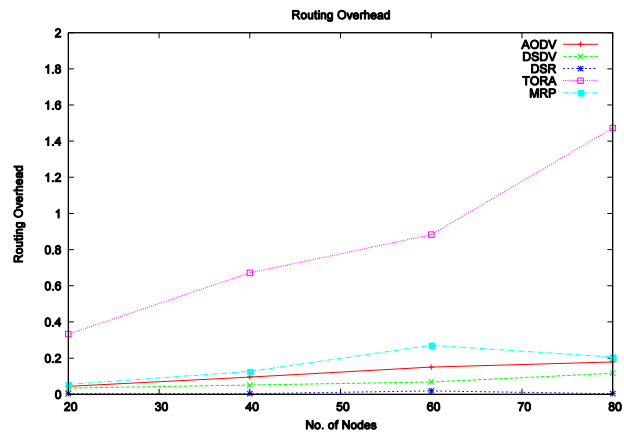

Figure 7: RO

In the above figure 7, it is clear that our proposed frame gives performance nearly equal to DSR. In other words MRP gives best performance in ROH.

## V. CONCLUSION

This research paper I give a summary of Wireless Ad-hoc network and talk about that how these types of networks requires performance as most essential constraints. A comprehensive study of the performance analysis strategies and performance analysis metrics is provided. According to this study it is focusing on four performance analysis methods to achieve better performance. I evaluate and simulate four routing protocols with my proposed MRP framework to investigate the performance analysis of routing protocols. On the basis of my simulation results it is clear that my proposed framework to increase performance of routing protocols. The idea of this research work is to develop an efficient performance routing protocol and allows researchers to select the well define routing method.

**AUTHOR'S PROFILES**

Sohail Abid: (Mobile No: +92-321-5248497)
Sohail Abid MS in Telecomm and Network and working as System Administrator at Foundation University Rawalpindi Campus.

Shahab Khan: (Mobile No: +92-301-8531492)
Shahab Khan Student of MS (TN) at IQRA University Islamabad.